\documentclass[3p,times,twocolumn,longtitle]{elsarticle}
\usepackage{ecrc}

\volume{00}

\firstpage{1}

\journalname{Nuclear Physics B Proceedings Supplement}

\runauth{M. Sisti et al.}

\jid{nuphbp}

\jnltitlelogo{Nuclear Physics B Proceedings Supplement}

\usepackage{amssymb}

\biboptions{compress}

\usepackage[figuresright]{rotating}

\def\teutz{$^{130}$Te}

\def\biduq{$^{214}$Bi}
\def\tldzo{$^{208}$Tl}

\def\udto{$^{238}$U}
\def\thdtd{$^{232}$Th}
\def\kqz{$^{40}$K}
\def\cosz{$^{60}$Co}
\def\rnddd{$^{222}$Rn}
\def\ptunz{$^{190}$Pt}

\def\teod{TeO$_2$}
\def\teodnat{$^{\mathrm{nat}}$TeO$_2$}

\def\bbzn{$\beta\beta0\nu$}
\def\bbdn{$\beta\beta2\nu$}
\def\mbb{$|\langle m_{\beta \beta}\rangle |$}
\def\me{$m_e$}
\def\Qbb{$Q_{\beta\beta}$}

\def\rate{counts/(keV$\cdot$kg$\cdot$y)}
\def\vita-dim{$\tau_{1/2}$}

\begin{document}

\begin{frontmatter}



\dochead{}

\title{Status of the CUORE and results from the CUORE-0 neutrinoless
   double beta decay experiments}
%
%
\author[Milano,INFNMiB]{M.~Sisti\corref{cor1}}
\ead{monica.sisti@mib.infn.it}
\author[USC,LNGS]{D.~R.~Artusa}
\author[USC]{F.~T.~Avignone~III}
\author[Legnaro]{O.~Azzolini}
\author[LNGS]{M.~Balata}
\author[BerkeleyPhys,LBNLNucSci,LNGS]{T.~I.~Banks}
\author[INFNBologna]{G.~Bari}
\author[LBNLMatSci]{J.~Beeman}
\author[Roma,INFNRoma]{F.~Bellini}
\author[INFNGenova]{A.~Bersani}
% ...

\address[Milano]{Dipartimento di Fisica, Universit\`{a} di Milano-Bicocca, Milano I-20126 - Italy}
\address[INFNMiB]{INFN - Sezione di Milano Bicocca, Milano I-20126 - Italy}
\address[USC]{Department of Physics  and Astronomy, University of South Carolina, Columbia, SC 29208 - USA}
\address[Legnaro]{INFN - Laboratori Nazionali di Legnaro, Legnaro (Padova) I-35020 - Italy}
\address[LNGS]{INFN - Laboratori Nazionali del Gran Sasso, Assergi (L'Aquila) I-67010 - Italy}
\address[BerkeleyPhys]{Department of Physics, University of California, Berkeley, CA 94720 - USA}
\address[LBNLNucSci]{Nuclear Science Division, Lawrence Berkeley National Laboratory, Berkeley, CA 94720 - USA}
\address[INFNBologna]{INFN - Sezione di Bologna, Bologna I-40127 - Italy}
\address[LBNLMatSci]{Materials Science Division, Lawrence Berkeley National Laboratory, Berkeley, CA 94720 - USA}
\address[Roma]{Dipartimento di Fisica, Sapienza Universit\`{a} di Roma, Roma I-00185 - Italy}
\address[INFNRoma]{INFN - Sezione di Roma, Roma I-00185 - Italy}
% ...
\address[INFNGenova]{INFN - Sezione di Genova, Genova I-16146 - Italy}
% ...

\cortext[cor1]{Corresponding author}
\fntext[fn1]{Deceased}
%
\begin{abstract}
CUORE is a 741\,kg array of TeO$_2$ bolometers for the search of neutrinoless double beta decay of $^{130}$Te. The detector is being constructed at the Laboratori Nazionali del Gran Sasso, Italy, where it will start taking data in 2015. If the target background of 0.01\,\rate\ will be reached, in five years of data taking CUORE will have a 1$\sigma$ half life sensitivity of 10$^{26}$\,y. CUORE-0 is a smaller experiment constructed to test and demonstrate the performances expected for CUORE. The detector is a single tower of 52 CUORE-like bolometers that started taking data in spring 2013. The status and perspectives of CUORE will be discussed, and the first CUORE-0 data will be presented.
\end{abstract}


\end{frontmatter}


\section{Introduction}
\label{sec:intro}

Neutrinos are massive particles. A beautiful proof of this important property was 
obtained by neutrino oscillation experiments more than a decade ago.
Since then, the key role of neutrinoless double beta decay searches has been established, as attested by the growing number of experimental proposals in the last years. 

Neutrinoless double beta decay (\bbzn) is a proposed very rare nuclear process in which a nucleus transforms into its (A,\,Z+2) isobar with the emission of two electrons. While the two neutrino channel (\bbdn) -- where two neutrinos are contemporary emitted in the decay -- is allowed by the Standard Model of Particle Physics and has been  observed experimentally in a
dozen of isotopes with half-lives of the order
$10^{18}-10^{21}$\,y, the neutrinoless mode is a lepton number violating process which can occur only if the Majorana character of the neutrino is allowed. 
Therefore, \bbzn\ offers a unique experimental chance to investigate still unresolved fundamental questions, since its observation would undoubtedly unveil the neutrino character, confirm lepton number violation and allow to assess the absolute neutrino mass scale with high sensitivity, thus helping point us towards the proper extension of the Standard Model \cite{bil12,ell02,avi08}.   

Neutrinoless double beta decay can proceed via different mechanisms: in the case of a
virtual exchange of a light Majorana neutrino between the
two nucleons, the decay rate is proportional to the square
of the so-called effective Majorana mass \mbb\ (a coherent sum of neutrino mass eigenstates) \cite{ber12}: 
\begin{equation}
\label{eq:half-life}
[T_{1/2}^{0\nu}]^{-1}=\frac{|\langle m_{\beta \beta }\rangle|^{2}}{m_{e}^2}G^{0\nu }|M^{0\nu }|^{2}
\end{equation}
where ${T}_{1/{2}}^{0\nu }$ is the decay half-life, ${G}^{0\nu }$ is the
two-body phase-space integral, ${M}^{0\nu }$ is the \bbzn\
Nuclear Matrix Element (NME), and  \me\
is the electron mass. The product $F_{N}^{0\nu }=G^{0\nu }|M^{0\nu}|^{2}$ includes all the nuclear details of the decay and it is usually
referred to as \textit{nuclear factor of merit}. While ${G}^{0\nu }$
can be calculated with reasonable accuracy, the NME value is strongly
dependent on the nuclear model used for its evaluation so that discrepancies of about a
factor 2-3 among the various theoretical calculations may be found \cite{suh98,sim99,kot12,men09,fae12,fan11,suh12,bar13,rat10,rod10}. Such uncertainties are of course  reflected on the \mbb\ inferred values.

Experimentally one measures the energy deposited by the two electrons, which result in a peak centered at the \bbzn\ candidate isotope transition energy (\Qbb). For neutrino masses in the Inverted Hierarchy region, half-lives in the range $10^{26}-10^{27}$\,y are expected for several \bbzn\ candidates \cite{ihe}. This implies a few decays per 100\,kg of candidate isotope per year. In a realistic experiment this faint signal must be singled out among the background events in the energy region around \Qbb. Hence, the sensitivity of a given experiment critically depends on the number of spurious counts in the region of interest: for a 68\% confidence level, it is defined as the decay half-life corresponding to the maximum signal that could be hidden by a 1$\sigma$ background fluctuation $n_B=\sqrt{BT\Delta M}$, where $\Delta$ is the FWHM energy resolution, $M$ is the detector mass, $T$ is the measuring time, and $B$ is the background per unit mass, energy, and time. The experimental figure of merit is thus given by:
\begin{equation}
\label{eq:sensitivity}
F^{0\nu}=T_{1/2}^{Back.Fluct.}=\frac{\ln 2 N_{\beta \beta } T}{n_B} \propto \eta \epsilon \sqrt{\frac{MT}{B\Delta}}
\end{equation}
 where $N_{\beta \beta }$ is the number of \bbzn\ decaying nuclei under observation, $\eta$ is the \bbzn\ candidate isotopic abundance and $\epsilon$ is the detection efficiency. This clearly shows that the sensitivity to the \bbzn\ signal goes linearly with the isotopic fraction and the detection efficiency, as the square root of mass and measuring time and, much worse -- given the relation between half-life and \mbb\ -- the improvements on the assessment of the neutrino mass have only a fourth root dependence on the experimental parameters.

\section{The CUORE experiment}
\label{sec:cuore}

\begin{figure}
\centering
\resizebox{0.3\textwidth}{!}{%
  \includegraphics{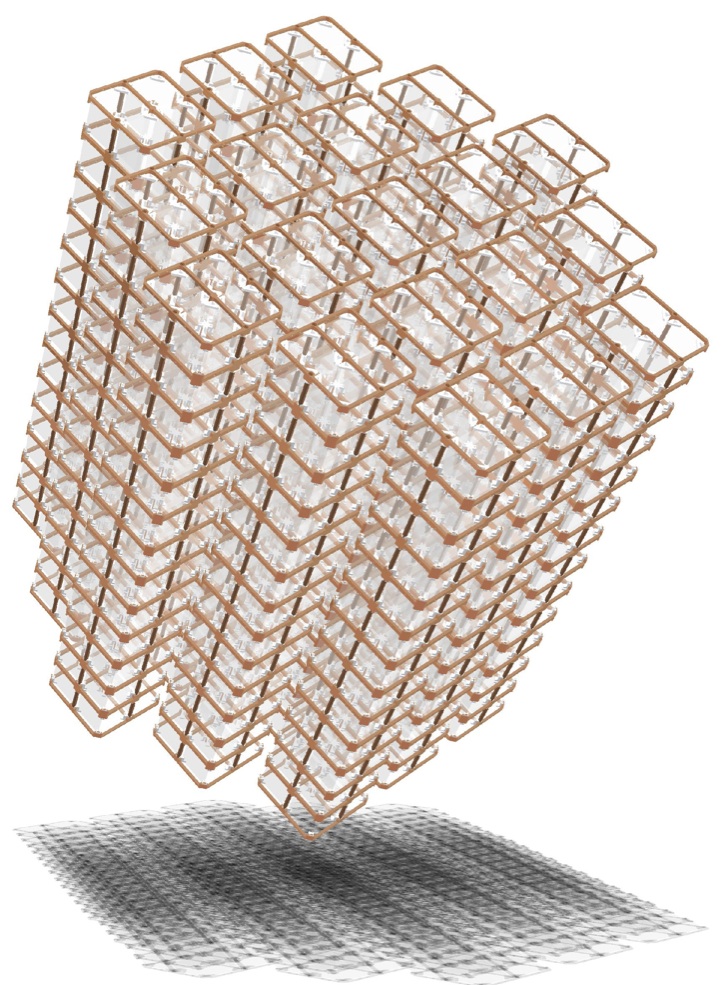}
}
\caption{The CUORE detector array scheme.}       
\label{fig:cuore-array}
\end{figure}

\begin{figure}
\centering
\resizebox{0.1\textwidth}{!}{%
  \includegraphics{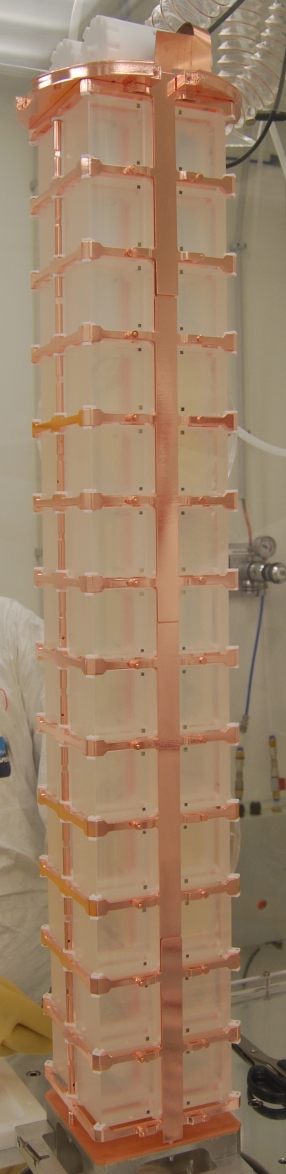}
}
\caption{The CUORE-0 tower before its installation in the cryostat.}       
\label{fig:cuore0}
\end{figure}

CUORE (Cryogenic Underground Observatory for Rare Events) is an array of 988 \teodnat\ low temperature detectors (or bolometers) with the primary goal of searching for the \bbzn\ of \teutz\ at Laboratori Nazionali del Gran Sasso (LNGS), Italy \cite{cuore-proposal}.
The array is a compact structure of 19 towers, each one containing 52 detectors arranged 
on 13 planes (see Figures \ref{fig:cuore-array} and \ref{fig:cuore0}). Every detector is a low temperature calorimeter, i.e. a very sensitive device operated at mK temperatures in which any energy release in the \textit{absorber} is measured via its temperature rise through a suitable thermal sensor (the \textit{thermometer}) \cite{fiorini-niini}. In principle any dielectric and diamagnetic crystal can act as a particle absorber, therefore a wide choice of materials is made available by such technology. This characteristics together with the excellent energy resolution make low temperature calorimetry an ideal technique for experiments in the rare event physics field. In particular, \teod\ crystals are very convenient low temperature absorbers because of their low specific heat, their intrinsic radiopurity and their crystallographic properties, which allow to grow excellent crystals of relatively large mass. 
Moreover, \teod\ contains the isotope \teutz, a quite appealing \bbzn\ candidate due to its large natural isotopic abundance (34\%), its high transition energy (\Qbb$=2528$\,keV, above most  $\gamma$ background lines from natural radioactivity), and its encouraging nuclear matrix element evaluations.
In the CUORE array each detector is composed by a $5 \times 5 \times 5$\,cm$^3$ \teod\ crystal (750\,g of mass) on whose surface a Neutron Transmutation Doped (NTD) Ge thermistor is glued to measure any temperature change of the absorber and convert it into an electric signal. An electro-thermal link restores the operating temperature after every interaction. A Joule heater is also glued on the surface of the \teod\ crystals to inject a known amount of energy in each detector for off-line thermal gain corrections (all detectors will be operated independently). 
The \teod\ crystals are secured in groups of four inside low background copper frames by means of Teflon supports; each of the 19 CUORE towers is composed by 13 such elements. The total mass of the CUORE array amounts to 741\,kg of \teod\ (206\,kg of \teutz).

The CUORE detectors will be installed in a complex cryogenic apparatus where they will be operated at a temperature of $\sim$\,10\,mK. The low temperature setup includes a custom cryostat (with six nested copper shields), a cryogen free cooling system with five pulse tubes, a very powerful dilution refrigerator, and a fast cooling system for pre-cooling. All materials used to build the experimental apparatus have been selected following stringent requirements concerning their radiopurity. The detector towers will be hanging on a thick copper disk thermally anchored to the 10\,mK stage and mechanically decoupled from the rest of the setup to minimize vibrational noise.  A set of cold roman lead \cite{pbrom} radioactive shields will screen the detectors from the radioactive contaminations of the cryogenic apparatus. A detector calibration system consisting of twelve thoriated wires, to be guided in proximity of the towers inside small copper tubes, has been designed to uniformily irradiate the detectors during the monthly energy calibrations with \thdtd\ $\gamma$ rays. Finally, heavy room temperature shields made by layers of borated polyethylene, boric-acid powder, and lead bricks surround the apparatus to reduce the background contribution from neutrons and $\gamma$ rays. 
The CUORE experimental site is located in Hall A of LNGS, under a rock overburden of $\sim$\,3600\,m of water equivalent, which reduces the muon flux to $\sim$\,3$\times 10^{-8}$\,$\mu$/s/cm$^2$ \cite{mei06}. Presently CUORE is in the final phase of its construction (see Section\,\ref{cuore-status})  and it is expected to start taking data in 2015.

\teod\ low temperature detectors (or bolometers) have a long history in \bbzn\ searches of \teutz\ \cite{arn-PRC78, and-AP34}, the best sensitivity being achieved by Cuoricino, a $\sim$\,40\,kg tower of 62 bolometers ($\sim$\,11\,kg of \teutz) operated at LNGS site in the years 2003--2008: after an exposure of 19.75\,kg\,y of \teutz, Cuoricino set a lower limit on the decay half-life of this isotope of $2.8 \times 10^{24}$\,y (90\% C.L.) \cite{and-AP34}. 
To improve that result, as already discussed (see \ref{eq:sensitivity}), one needs to increase the experimental mass, improve  the energy resolution and, what's even more tricky, reduce the radioactive background contribution in the energy region of interest (ROI -- centered on \Qbb). 

CUORE design goals, besides the increase of the detector mass of a factor of $\sim$\,20 with respect to Cuoricino, are a FWHM resolution of 5\,keV at an energy around 2.5\,MeV, and a background rate in the ROI of $10^{-2}$\,\rate. This translates in a 1$\sigma$ (90\% C.L.) sensitivity of $1.6 \times 10^{26}$\,y ($9.5 \times 10^{25}$\,y) in 5\,y of measuring time, corresponding to an effective Majorana mass sensitivity in the range 50\,--\,130\,meV.

\section{The CUORE background reduction strategy}
\label{sec:background}

Cuoricino final sensitivity was limited by the background rate in the ROI, which was dominated by the surface radioactive contaminations of the detector materials -- \teod\ crystals and copper structures.
Therefore, in preparing the CUORE array, several improvements have been implemented.
First of all the detector design has been conceived to minimize the amount of passive materials close to the \teod\ crystals: only selected radiopure copper and Teflon have been used for the tower skeleton and the total mass of the copper frame was reduced  by a factor of 2.3 with respect to Cuoricino.  Besides decreasing the copper surface background contribution to the ROI, this fact helps also optimizing the time coincidence analysis between the detectors (unlike background interactions, a \bbzn\ event is fully contained inside a single $5 \times 5 \times 5$\,cm$^3$ \teod\ crystal, with a calculated efficiency of 87\%).
To further mitigate the $\alpha$ background component a specially developed surface treatment protocol was applied to all copper parts directly facing the detectors. This procedure, called TECM (for Tumbling, Electropolishing, Chemical etching, and Magnetron plasma etching) has demonstrated to efficiently reduce the concentration of both \udto\ and \thdtd\ on the copper surface to better than $1.3 \times 10^{-7}$\,Bq/cm$^2$ \cite{TTT}. Moreover, to limit cosmogenic activation, all the copper material of the CUORE experiment has been stored underground at LNGS in between production steps.\\
Finally, a strict production protocol to limit bulk and surface contamination of the \teod\ crystals has been developed in collaboration with the crystal manufacturer at the Shangai Institute of Ceramics, Chinese Academy of Science \cite{crescita-cristalli}. After production, the \teod\ crystals were transported to LNGS at sea level to minimize cosmogenic activation. Cryogenic tests on few crystals from different production batches demonstrated bulk and surface contamination levels of $6.7 \times 10^{-7}$\,Bq/kg and $8.9 \times 10^{-9}$\,Bq/cm$^2$ at 90\%\,C.L., respectively, for \udto,  and  of $8.4 \times 10^{-7}$\,Bq/kg and $2.0 \times 10^{-9}$\,Bq/cm$^2$ at 90\%\,C.L., respectively, for \thdtd\ \cite{ccvr}.

Special care was also devoted to the CUORE assembly line. A dedicated class 1000 clean room area was constructed at the first floor of the underground CUORE experimental building and is equipped with a set of specially designed glove boxes, inside which all operations needed to assemble a detector tower are performed under radon-free atmosphere. 
All materials used to construct the glove boxes were carefully selected, and all tools used inside the glove boxes, especially those that come into physical contact with the detector components, were cleaned and certified for their radiopurity.
The clean area is divided into four rooms, each one devoted to specific tasks. 
In the gluing area thermistors and heaters are attached to the crystals by means of a semi-automated robotic system, to guarantee the uniformity and repeatability of the glue joints. In the assembly area, the instrumented crystals are secured inside the copper frames with the teflon spacers and then mechanically assembled into a tower. After that, a set of flexible printed circuit board cables is attached to the tower and \textit{in situ} wire bonding is performed for the electrical connections of the detectors.
The completed towers are then put up in the storage area inside specially designed nitrogen-flushed boxes to await the installation in the cryostat. The fourth room of the clean area provides direct access to the experimental area of the cryogenic setup, so as to always preserve the detectors in a protected environment.
More details on the CUORE detector assembly can be found in \cite{ahep-2014}.

In order to check the effectiveness of all the actions undertaken to improve Cuoricino results, we decided to run the CUORE-0 experiment. CUORE-0 is one CUORE-like tower built using the same detector components, cleaning protocols and assembly procedures defined for CUORE. CUORE-0 is therefore a proof of concept of the CUORE detector at all stages, including data acquisition and data analysis frameworks. Besides that, CUORE-0 is an independent \bbzn\ experiment on its own, which should extend the physics reach beyond Cuoricino while CUORE is being assembled.

\section{CUORE-0}
\label{sec:cuore-0}

CUORE-0 is an array of 52 \teodnat\ detectors of 750\,g each, for a total mass of 39\,kg ($\sim$11\,kg of \teutz). 
It is operated in the same experimental setup that hosted Cuoricino, i.e. a 25 year-old cryostat located in Hall A of LNGS, with an external shielding made of lead and borated-polyethylene, and enclosed inside a Faraday cage \cite{arn-PLB557}. Also the front-end electronics and data acquisition hardware are the same used for Cuoricino.  The waveforms coming from the detectors are amplified, filtered with a six-pole active Bessel filter, and then fed into a National Instrument PXI analog-to-digital converter (ADC) with 18-bit resolution and 125\,S/s sampling frequency. All samples are continuously saved to disk and  then processed with Apollo, the data acquisition software developed for CUORE, which identifies pulses with a constant fraction software trigger.     
In addition, a heater pulse is injected every 300\,s to each bolometer for stabilization monitoring, and baseline samples are recorded at 200\,s intervals to keep under control the working temperature and noise conditions of each detector. 
The signals have a typical time evolution of $\sim$5\,s, with rise times in the range 40\,--\,80\,ms and decay times in the range 100\,--\,700\,ms.\\
The energy calibration of the detectors is obtained by inserting two thoriated tungsten wires in between the external shielding and the cryostat. Usually, 2-3 days of acquisition time are enough to independently calibrate  each detector in the energy region between 511 and 2615\, keV with \thdtd\ $\gamma$ rays. Physics (or low background) runs are usually 3-4 weeks long. A CUORE-0 \textit{data set} is composed by an initial calibration, a low background run, and a final calibration.
The raw data are then processed offline with Diana, the analysis software developed for CUORE. The analysis procedure is basically the same applied to the Cuoricino data \cite{and-AP34}, the main steps being amplitude evaluation, gain correction, energy calibration, and time coincidence analysis among the detectors. Especially this last point serves as \textit{active background rejection} technique, which will be even more effective with the CUORE geometry: if any two or more detectors register signal pulses within 100\,ms from each other, the events are tagged as coincident and discarded. In fact the coincidence events are mostly attributed to background interactions, such as Compton-scattered $\gamma$ rays or $\alpha$ decays on the surface of nearby crystals, since the majority (87.4$\pm$1.1\% \cite{and-AP34}) of the \bbzn\ decay electrons are expected to be fully contained in a single crystal. 
Events are selected also on the basis of their shape: noisy pulses are discarded as a result of a comparison with a set of template signals.
The last step is the production of the total energy spectra (calibration and background) by summing all the single detectors spectra for the entire acquired statistics.
\begin{figure*}
\centering
\resizebox{1.0\textwidth}{!}{%
  \includegraphics{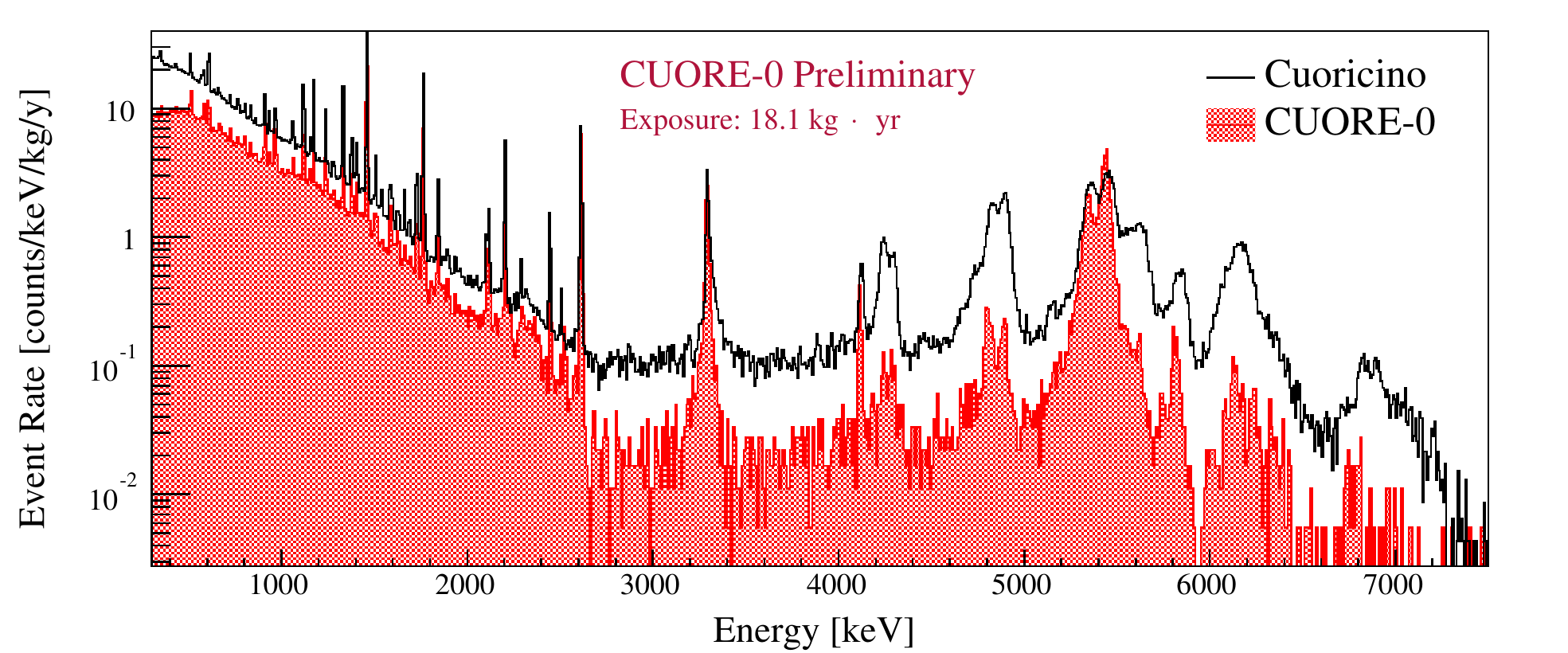}
}
\caption{Cuoricino (line) and CUORE-0 (shaded) background spectra comparison. Only events with a single crystal hit are considered (anti-coincidence cut). The strong background reduction achieved with the CUORE-0 detector is evident. It amounts to a factor of $\sim$6 in the $\alpha$ region (i.e. energies above 2.7\,Mev) and to a factor of $\sim$2 in the $\gamma$ region (i.e. energies below 2.7\,Mev).}       
\label{fig:cuore0-fondo}
\end{figure*}

\begin{figure}
\centering
\resizebox{0.5\textwidth}{!}{%
  \includegraphics{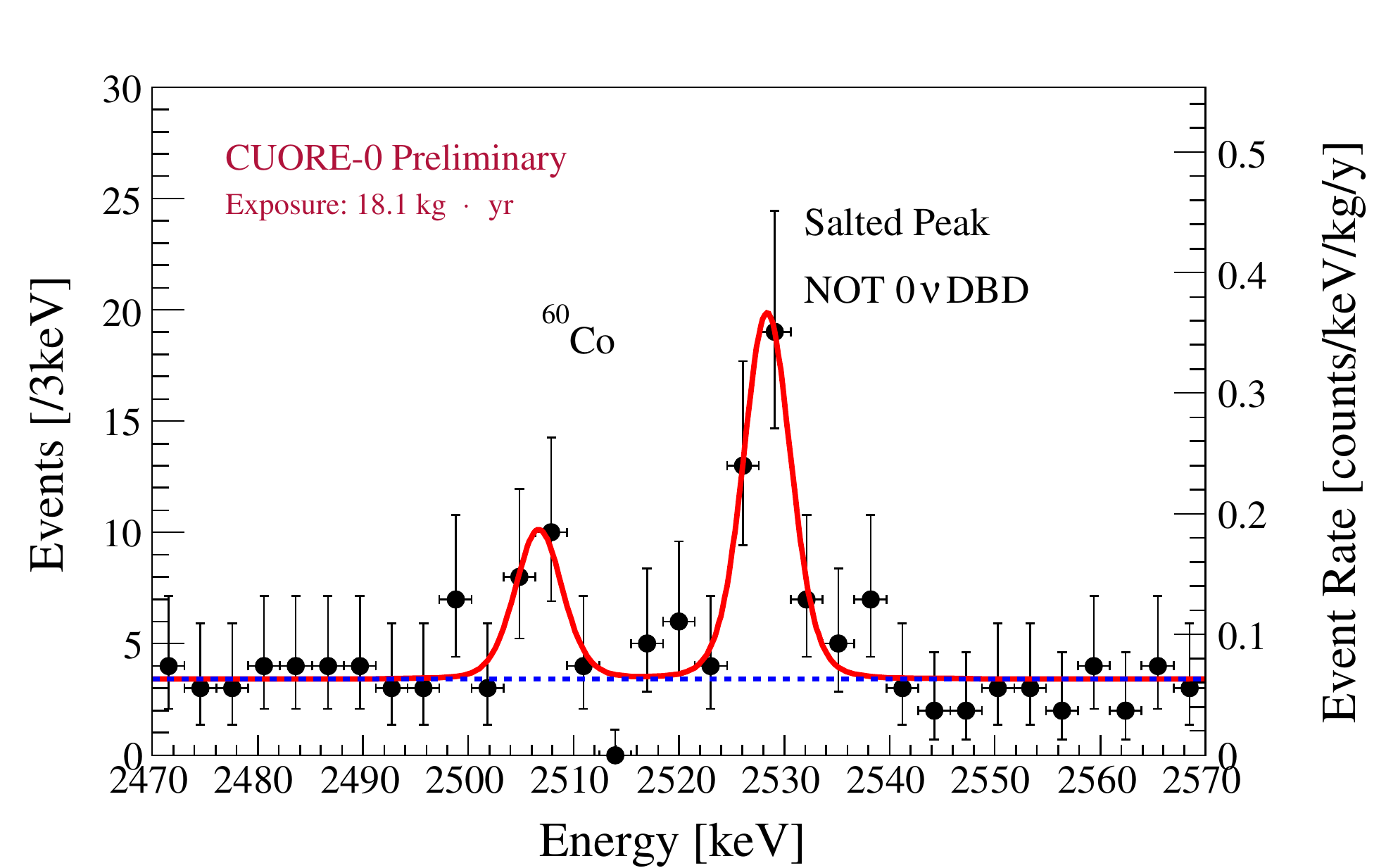}
}
\caption{The CUORE-0 \bbzn\ energy region of interest (ROI). The line at 2.505\,MeV is the \cosz\ sum peak while the one at 2.528\,MeV is the salted peak (see text).}       
\label{fig:cuore0-ROI}
\end{figure}

CUORE-0 data taking started in March 2013, though in not ideal conditions: the first results (Phase I) were published in summer 2013 \cite{vignati-taup13,cuore0-initial}. Then a long stop was needed for cryostat maintainance. The data taking resumed in November 2013, with strongly improved performance, and is still going on (Phase II). The total  spectrum obtained by summing all calibration spectra of all active channels (49 out of 52) acquired during Phase II shows a FWHM resolution at 2.615\,MeV (\tldzo\ line) equal to 4.8\,keV, better than aimed for CUORE (see Section\,\ref{sec:cuore}).
The total background exposure sums up to 18.1\,kg$\cdot$y (6.2\,kg$\cdot$y of \teutz) and covers the period from March 2013 until May 2014 (Phase I + Phase II). The overall detection efficiency is 
77.6\,$\pm$\,1.3\%, and includes the anti-coincidence cut, the pulse selection, and the containment efficiency.
The resulting spectrum is shown in Figure\,\ref{fig:cuore0-fondo}. 
It is possible to recognize the $\gamma$ lines of \tldzo, \kqz\ and \cosz, attributed to a \thdtd\ contamination of the cryostat, and the ones of \biduq, attributed to the presence of \rnddd\ in the air around the cryostat during the initial runs. For what concerns the peaks in the $\alpha$ region of the background spectrum, the main sources are bulk contaminations of the \teod\ crystals, while other structures are attributed to surface contaminations of the \teod\ crystals or of the passive materials close to them. The peak at $\sim$3.2\,MeV is attributed to \ptunz, a bulk contamination of the \teod\ crystals presumably originating by the platinum crucible during crystal growth. The continuum between 2.7\,MeV and 3.9\,MeV, excluding the \ptunz\ peak, is attributed to $\alpha$ particles of degraded energy, i.e. particles that deposit a fraction of their energy in a crystal and the rest in an inactive material, like the copper structure (therefore no coincidence cut is possible). This continuum extends down into the gamma region and in Cuoricino was responsible for a significant fraction ($\sim$50\,$\pm$\,20\%) of the background in the ROI. 
Figure\,\ref{fig:cuore0-fondo} shows the tremendous improvement achieved with CUORE-0: the background in the $\alpha$ continuum now amounts to 0.020\,$\pm$\,0.001\,\rate\ -- was 0.110\,$\pm$\,0.001\,\rate\  in Cuoricino -- a factor $\sim$6 improvement thanks to the background reduction strategy undertaken for CUORE. This translates in a background reduction in the ROI of a factor of $\sim$2.5 with respect to Cuoricino: in this case the count rate still accounts for a significant contribution from the environmental $\gamma$s coming from the old cryostat bulk contaminations, consistent with what was observed in Cuoricino. A closer look to the energy spectrum in the ROI is shown in Figure\,\ref{fig:cuore0-ROI}: the backgound rate is 0.063\,$\pm$\,0.006\,\rate\ -- was 0.153\,$\pm$\,0.006\,\rate\  in Cuoricino. The peak at the \bbzn\ energy is a false one intentionally produced by our blinding procedure: a small blinded fraction of the events within $\pm 10$\,keV of the \tldzo\ 2.615\,MeV background line is exchanged with the events within $\pm 10$\,keV of the \teutz\ \Qbb\ value. The fit function (solid line of Fig.\,\ref{fig:cuore0-ROI}) is a flat background with two gaussians, one for the \cosz\ 2.505\,MeV sum peak and one for the salted peak.
The data unblinding is expected in Spring 2015, when the CUORE-0 sensitivity should surpass the Cuoricino one.
CUORE-0 data taking will instead continue until CUORE starts.

\section{CUORE status and prospects}
\label{cuore-status}

CUORE is currently at an advanced state of construction. The detector is fully assembled: the 19 towers are ready and safely stored inside nitrogen flushed canisters, while waiting to be installed inside the cryostat.

Major efforts are now devoted to the last phases of the cryostat preparation. Given the complexity of the system a phased installation was chosen, with a limited number of items being added at each step.\\
The commissioning plan started in 2012. Currently, the cryogenic system is fully assembled and the first successfull cooldown was recently completed, showing a base temperature ($\sim$7\,mK) safely below the design value of 10\,mK. Next steps foresee the installation of the calibration system, of the wiring system, of the cold shields, and of the detector suspension plate. 
The final step will be the tower installation. 

The CUORE experiment data taking is expected to start at the end of 2015. The encouraging results obtained with CUORE-0 seem to indicate that the CUORE design goal of 0.01\,\rate\ is within reach.

\section{Acknowledgments}
The CUORE Collaboration thanks the directors and staff of the
Laboratori Nazionali del Gran Sasso and the technical staff of our
laboratories. This work was supported by the Istituto Nazionale di
Fisica Nucleare (INFN); the National Science
Foundation under Grant Nos. NSF-PHY-0605119, NSF-PHY-0500337,
NSF-PHY-0855314, NSF-PHY-0902171, and NSF-PHY-0969852; the Alfred
P. Sloan Foundation; the University of Wisconsin Foundation; and Yale
University. This material is also based upon work supported  
by the US Department of Energy (DOE) Office of Science under Contract Nos. DE-AC02-05CH11231 and
DE-AC52-07NA27344; and by the DOE Office of Science, Office of Nuclear Physics under Contract Nos. DE-FG02-08ER41551 and DEFG03-00ER41138.
This research used resources of the National Energy Research Scientific Computing Center (NERSC).




\nocite{*}
\bibliographystyle{elsarticle-num}
\bibliography{martin}







\end{document}